\documentclass[aps,floatfix,prl,letterpaper]{revtex4}
\usepackage{graphicx}% Include n files
\usepackage{dcolumn}% Align table columns on decimal point
\usepackage{bm}% bold math
\usepackage{natbib}
\usepackage{subfigure}
\usepackage{epstopdf}
\usepackage{natbib}
\usepackage{amsmath} 
\usepackage{color} 
\usepackage{longtable} 
\usepackage{tabularx}
\usepackage[abs]{overpic}

\begin{document} 
\title{Elastohydrodynamics and kinetics of protein patterning in the immunological synapse} 
\author{A. Carlson}
\author{L. Mahadevan}
\email{lm@seas.harvard.edu}
\date{\today}

\affiliation{School of Engineering and Applied Sciences, Kavli Institute for Bionano Science and Technology, and Wyss Institute, Harvard University, Cambridge, USA}
\affiliation{Departments of Physics, and Organismic and Evolutionary Biology, Harvard University, Cambridge, USA.}

\begin{abstract}%
{The cellular basis for the adaptive immune response during antigen recognition relies on a specialized protein interface known as the immunological synapse (IS). Understanding the biophysical basis for protein patterning by deciphering the quantitative rules for their formation and motion is an important aspect of characterizing immune cell recognition and thence the rules for immune system activation. We propose a minimal mathematical model for the physical basis of membrane protein patterning in the IS, which encompass membrane mechanics, protein binding kinetics and motion, and fluid flow in the synaptic cleft. Our theory leads to simple predictions for the spatial and temporal scales of protein cluster formation, growth and arrest as a function of membrane stiffness, rigidity and kinetics of the adhesive proteins, and the fluid in the synaptic cleft. Numerical simulations complement these scaling laws by quantifying the nucleation, growth and stabilization of proteins domains on the size of the cell. Direct comparison with experiment shows that passive elastohydrodynamics and kinetics of protein binding in the synaptic cleft can describe the short-time formation and organization of protein clusters, without evoking any active processes in the cytoskeleton.  Despite the apparent complexity of the process, our analysis highlights the role of just two dimensionless parameters that characterize the spatial and temporal evolution of the protein pattern: a ratio of membrane elasticity to protein elasticity, and the ratio of a hydrodynamic time scale for fluid flow relative to the protein binding rate, and we present a simple phase diagram that encompasses the variety of patterns that can arise. }%1
\end{abstract}
\maketitle %%The above information typeset through this command

\section{introduction}
Recognition of self or non-self is essential for an effective and functional adaptive immune response. The main players in this process are immune cells (T-lymphocyte cells (T-cells) \cite{Lanzavecchia:1985,Monks:1998, Grakoui:science1999}, B-cells, natural killer (NK) cells \cite{Davis:1999} and phagocytes \cite{Stinchcombe:Nat2006,Ravi:2007} that are constantly on the move scanning surfaces for antigenic peptides on Antigen Presenting Cells (APC). Receptors on the membrane of the immune cells are responsible for sensing and translating information from the extracellular matrix into the cell. Upon antigen recognition the immune cell  orchestrates a spatio-temporal organization of its membrane bound proteins that form a protein interface, known as the Immunological Synapse (IS) \cite{Norcross:1984}. More broadly, intercellular signaling in a functional IS relates to the formation of large protein domains \cite{Monks:1998, Grakoui:science1999}, whereas the time scale for their formation and the cluster-to-cluster interaction plays an important role in determining the overall cell signaling mechanism. This in turn depends on characterizing the dynamics of the pattern itself, which requires us to consider cellular membrane deformations by the receptor-ligand interaction and active cytoskeleton processes, leading to fluid motion in the synaptic cleft and thence IS dynamics \cite{BemillerFIM2013}. 

In the widely studied T-cells, the compartmentalization of membrane-bound protein patterns into different protein domains on the cellular scale leads to the formation of Supra Molecular Activation Clusters (SMACs) \cite{Monks:1998, Grakoui:science1999}. In particular, T-Cell Receptors (TCR) form bonds with the peptide Molecular HistoComplex (pMHC) on the APC, while Leukocyte-Function-Associated antigen-1 (LFA)-integrin on the T-cell bind with Intercellular Adhesion Molecules (ICAM) \cite{Grakoui:science1999}. {A few seconds ($O(1~s)$) \cite{Campi:2005} after membrane-to-membrane contact sub micron protein clusters are formed that start to translocate ($O(1~min)$) \cite{Grakoui:science1999}. This is followed by long range transport and a concomitant coarsening to form large-scale protein domains at longer times ($O(40~min)$) \cite{Grakoui:science1999,Varma:2006Imm}}. Observations of the T-cell IS show a central accumulation of TCR-pMHC, surrounded by a donut-shaped preferential protein domain of LFA-ICAM \cite{Monks:1998}.  

Understanding the biophysical basis for protein patterning by deciphering the quantitative rules for their formation and motion \cite{Beemiller:2012} is a first step in characterizing recognition and communication in the immune system. A particularly interesting question in this regard is the role of passive physicochemical processes relative to active motor-driven processes in generating these patterns \cite{James:2012Nat}. {Recent experiments suggest that early on during the process, active processes may not be important, and it is only later that the protein pattern in the T-cell membrane is subject to cytoskeletally generated centripetal transport  \cite{Ilani:2009NatIm, yi:2012, Babich:2012,Kaizuka:2007,Hartman:PNAS09,hammer:2013,Mossman:2005}.  The question of characterizing the mechanics of the IS patterns has led to range of mathematical models that take one of two forms: those that treat the system as a collection of discrete units \cite{Weikl:EPL2002, Weikl:2004Biophys,Figge:epjD2009,Paszek:2009,Reister:NJP2011} or as a continuum \cite{Qi:PNAS2001,Allard:BioP2012,BurroughsMerwe:2007}. While these models are capable of explaining the spatial patterning seen in the IS, they all neglect the fluid flow in the synaptic cleft and thus rely on ad-hoc assumptions for the characteristic time scales over which the patterns form, and use approaches based on gradient descent \cite{Qi:PNAS2001,Allard:BioP2012,BurroughsMerwe:2007,Burroughs:BioP2002} or stochastic variations of energy minimization of the membrane coupled to protein kinetics \cite{Weikl:EPL2002, Weikl:2004Biophys,Figge:epjD2009,Paszek:2009}.  

Here, we provide a description of the passive responses in the IS which includes the mechanical forces due to stretching and bending of the cell membrane which are driven by protein attachment and fluid flow, which itself causes flow of the trans-membrane proteins. This requires that we integrate cell membrane bending and tension, viscous flow in the membrane gap and protein attachment-detachment kinetics, and allows us to capture the essential spatiotemporal protein dynamics (nucleation, translation and coalescence of protein clusters) during the formation of SMACs.  Furthermore, we show that our description of the passive dynamics in the IS implies that the slow dynamics of fluid flow can limit the rate of protein patterning, without evoking any active cytoskeletal processes.} 

%\section{methods}
\section{Mathematical model}
\subsection{Membrane mechanics}
\begin{figure}
\center
\subfigure[{Sketch of the interaction between a T-lymphocyte cell (T-cell) and an glass supported antigen seeded bilayer.}]{%$(\frac{h_0}{L})^2\times t=7.0$]{
{\includegraphics[width=0.70\linewidth]{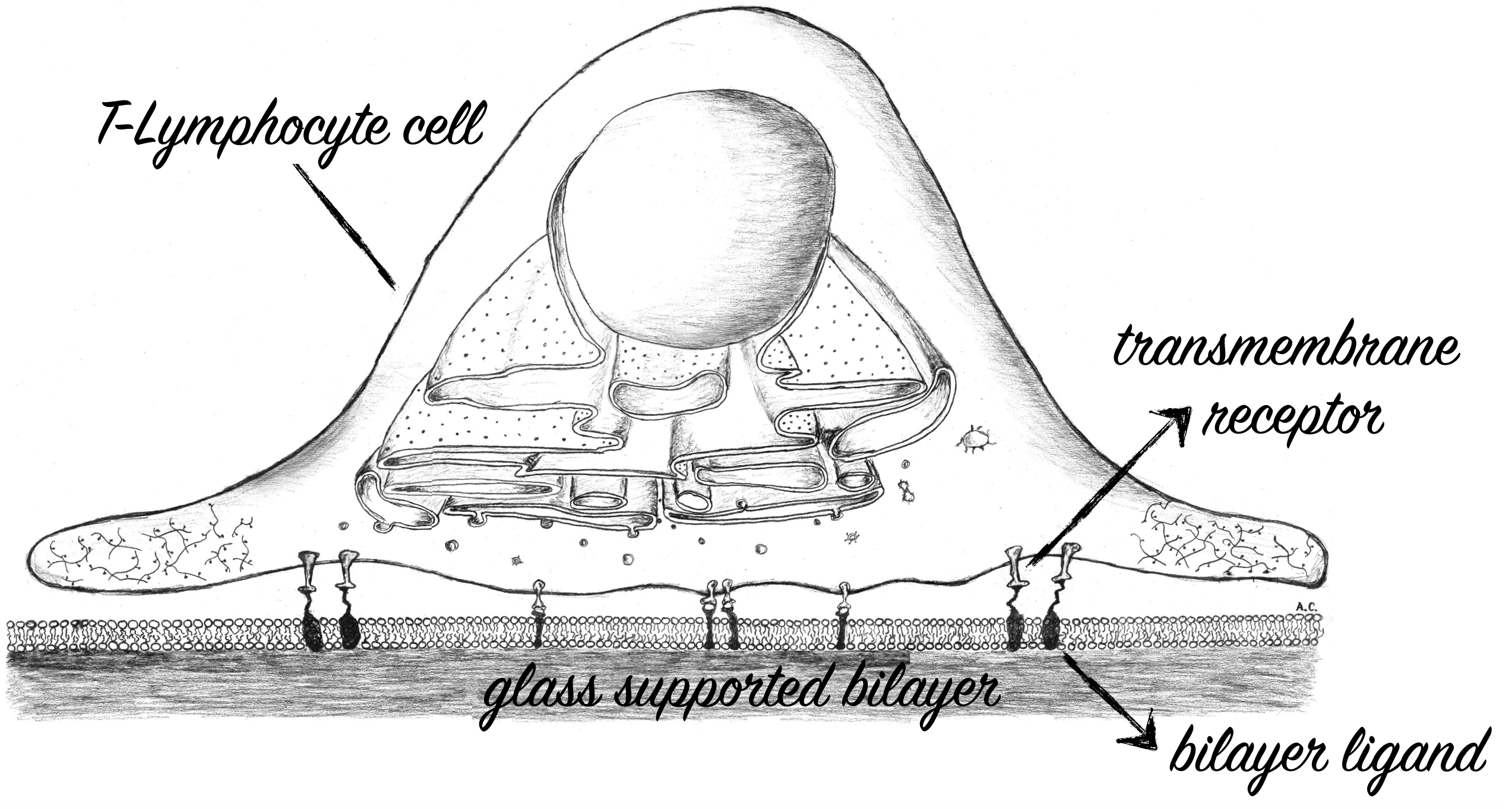}}
}
\subfigure[{Close view of the synaptic cleft formed between the T-cell membrane and the glass supported bilayer.}]{%$(\frac{h_0}{L})^2\times t=14.0$]{
{\includegraphics[width=1.0\linewidth]{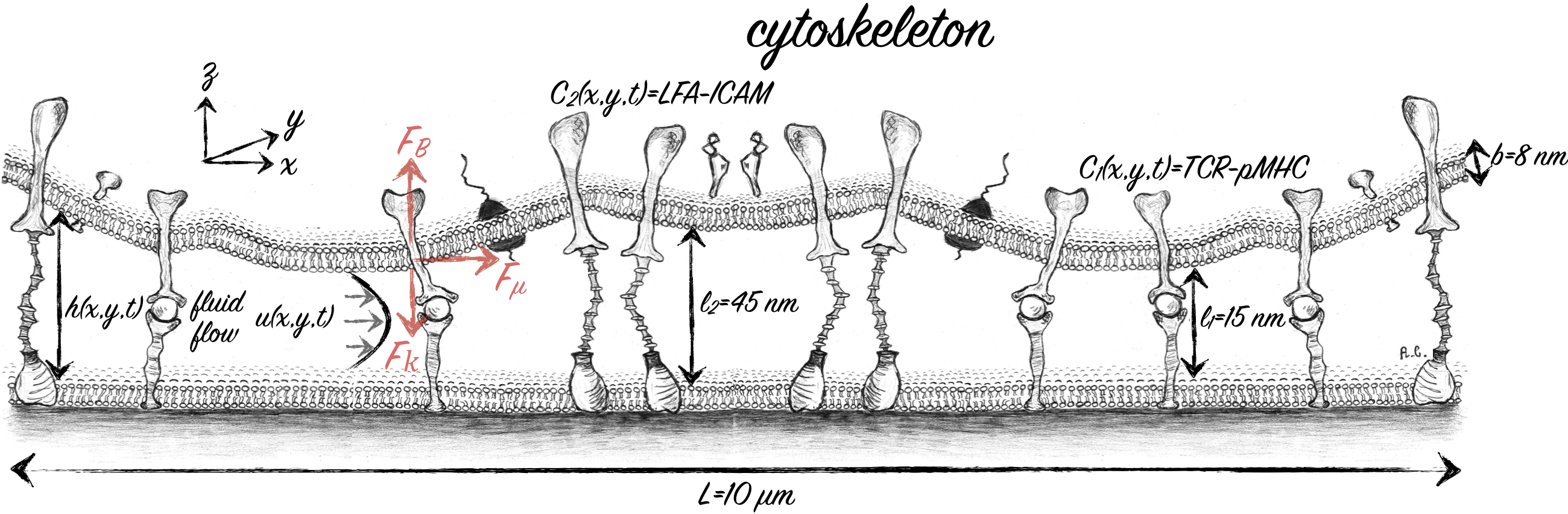}}}
\caption{(a) Sketch of the interaction between a T-lymphocyte cell (T-cell) and a supported antigen-seeded bilayer. The cell size $L\sim 10~\mu m$ and the two membranes are separated by transmembrane receptors bound to ligands in the bilayer. (b) Close-up schematic view of the synaptic cleft formed between the T-cell membrane and the glass supported bilayer. The cell membrane has a thickness $b\approx 8~nm$ and the membrane gap height is given by $h=h(x,y,t)$. The trans-membrane receptors form bonds with the ligands in the bilayer and have lengths and concentrations, $TCR-pMHC\approx 15~nm$, $C_1(x,y,t)$, and $LFA-ICAM\approx 45~nm$, $C_2(x,y,t)$. During protein bond formation and depletion, the cell membrane deforms generating a viscous flow $\mathbf{u}(x,y,z)$ in the membrane gap. The flow generates a viscous frictional force $F_{\mu}$ parallel with the glass supported bilayer that acts onto the cell membrane and the transmembrane proteins and thus affects their motion. Any deformation of the membrane generates a restoring elastic bending force $F_B$, while the deformation of the TCR-pMHC and LFA-ICAM bonds generates a  spring force $F_{\kappa}$. \label{fig:fig1}}
\end{figure}

In Fig. \ref{fig:fig1} we illustrate the interaction between a T-cell and an antigen seeded bilayer, which mimics the most commonly used experimental setup \cite{Grakoui:science1999,Kaizuka:2007,Hartman:PNAS09,Mossman:2005}, and describes the components in the mathematical model described below. Once the T-cell is close to the bilayer (Fig. 1) the membrane-bound receptors form adhesive bonds with their ligand counterparts in the bilayer, which pull the membranes together and squeezing the fluid out of the cleft.

When these two types of receptors form bonds with ligands, they get compressed or stretched. We assume that their spring stiffnesses $\kappa_i=\frac{l_1}{l_i}\kappa$ are inversely proportional to the protein length $l_i$ that may vary among different protein types \cite{Salas:2004}. The subscript $i=1$ corresponds to the TCR-pMHC complex and $i=2$ corresponds to the LFA-ICAM complex. $C_i=C_i(x,y,t)$ is the number of attached proteins per surface area (associated with at the total equilibrium receptor density $C_0$), their deformation creates a local pressure $\sim C_i(x,y,t)\kappa_i(l_i-h)$. This pressure deforms  the cell membrane, approximated here as a bilayer with a bending stiffness $B_m=\frac{Eb^3}{12(1-\nu^2)}$, with $E$ the Young's modulus, $b$ the membrane thickness and $\nu$ the Poisson ratio (see Supplementary Information (SI)), and a mechanical response quantified by 
\begin{equation}
\label{eq:p}
p(x,y,t)={B_m} \nabla^4 h+\kappa C_1(h-{l_1}) + \kappa \frac{2l_1}{l_2} C_2(h-l_2)
\end{equation}
where $p$ is the pressure difference across the membrane, and $h=h(x,y,t)$ is the height of the fluid-filled  synaptic cleft. By scaling the membrane gap with the longest protein bond $l_2$, the lateral lengths with the cell size $L$ and $p$ with the spring pressure $C_0\kappa l_2$ yields two non-dimensional numbers; $B=\frac{Eb^3}{12(1-\nu^2)\kappa C_0 L^4}$ describes the relative importance of pressure generated by membrane bending and the protein spring pressure, and $l_1/l_2=15nm/45nm=1/3$ is the ratio of the natural length of the proteins. {We focus here on the limit when membrane bending dominates, but we show in the SI that the influence of membrane tension smooths some of the small scale pattern features. Active cytoskeletal forces would appear as additional source terms in $p$, but has been neglected below as we focus on the passive dynamics.}

\subsection{Hydrodynamics}
Any membrane deformation initiates fluid motion and give rise to hydrodynamic forces in the synaptic cleft, which consequently affects the membrane dynamics. In typical experiments, the synaptic pattern has a lateral size $L$ comparable to the cell size ($\approx 10~\mu m$), while the cleft has a height comparable to size of the longest protein bond ($l_2=45~nm$). Thus the aspect ratio of the IS is small $l_2/L\ll1$. When combined with the fact that at these small length scales, the flow in the synaptic cleft is viscously dominated, we may use lubrication theory \cite{Batchelor:1967} to simplify the equations governing fluid flow. Under the assumption of a local Poiseuille flow \cite{Batchelor:1967} in the membrane gap {assuming no-slip at both surfaces, where the bilayer of the upper cell membrane can deform but the bilayer on the supported glass plate is immobilized. This leads  to a single non-linear scalar partial differential equation for the thin film height $h(x,y,t)$ \cite{Oron:1997} similar to that used in other elastohydrodynamic phenomena \cite{Hosoi:2004,Mani:2012,Leong:2010PRE}
\begin{equation}
\label{eq:tf}
\begin{split}
&\frac{\partial h}{\partial t}=\nabla \cdot \left(\frac{{h}^3}{12\mu}\nabla p\right), {\rm i.e.}\\
&\frac{\partial h}{\partial t}=\nabla \cdot \left(\frac{h^3}{12\mu}\nabla \left({B_m} \nabla^4 h+\kappa C_1({l_1}-h) + \kappa \frac{l_1}{l_2} C_2 (l_2-h)\right)\right),
\end{split}
\end{equation} 
where Eq. \ref{eq:tf} follows by using Eq. \ref{eq:p} for pressure ($p$), where $\mu$ is the fluid viscosity. Given the lack of evidence for water permeation across the membrane we neglect this effect, as well as thermal fluctuations of the membrane since these will be strongly damped out by enthalpic protein binding. 
%{A glycocalyx of proteins might fill the membrane gap with a porous flow of the interstitial fluid. By assuming a Darcy flow along the lateral directions $\mathbf{u}\approx \frac{K_G}{\mu} \nabla p$, where $K_G~[m^2]$ is the glycocalyx permeability and if $K_G\approx h^2$ the relationship for the fluid flow in Eq. 2 is recovered. It is possible that water may permeate the cell membrane, an effect that can be included by assuming that the flow through the membrane is porous and for a Darcy flow assuming the cell has a constant internal pressure ($p_c$) the right-hand side becomes $\approx \frac{K}{\mu b^2} p-p_c$ where $K~[m^2]$ is the membrane permeability and $b~[m]$ is the cell-membrane thickness. There are no experimental reports, to the best of our knowledge, that show any influence of water permeation in the IS and this effect is neglect below.} 

\subsection{Protein kinetics}
%Although there are free proteins in the synaptic cleft and in the cytoplasm of the cell in addition to those that are membrane bound, we focus only on those in the supported bilayer that are available for binding with the transmembrane proteins, i.e. we ignore the adsorption and desorption of proteins onto the bilayers. 
%
%This is equivalent to stating that the number of membrane-bound proteins is large compared to the free proteins, so that 
%
We only follow the dynamics of the membrane-bound proteins that can bind and unbind from their complementary ligands, which is equivalent to stating that the number of these proteins involved in the binding kinetics is large compared to the free proteins in the cytoplasm. In the membrane we assume the total number of membrane-bound proteins per unit area to be constant and given by $C_{i,0}$, where $i=1$ is corresponding to TCR and $i=2$ is corresponding to LFA. Of these, the number density of bound receptors is denoted by $C_i(x,y,t)$, which can diffuse and get dragged due to the fluid flow in the membrane gap, or be actively transported by the cytoskeleton.} Their dynamics can be described mathematically by a reaction-convection-diffusion equation, which accounts for these effects in addition to the binding and detachment of proteins, and in dimensional form reads
\begin{equation}
\label{eq:kin}
%\begin{split}
\frac{\partial C_i}{\partial t} =\frac{hl_i}{\mu}\nabla P \cdot \nabla C_i+\nabla\cdot\left(D_i\nabla C_i+\frac{k_bTD_i}{\mu}(C_i \nabla h (h-l_i))\right)+(C_{i,0}-C_i) {K}^{on}(l_i) -C_i{K}^{off}(l_i).
%\end{split}
\end{equation}
The first term on the right side is an advective term due to the fluid flow in the synaptic cleft driven by local pressure gradients associated with membrane deformation. The flow generates a Stokesian drag on the proteins proportional to their size. The second term is a membrane protein flux due to molecular diffusion $D_i\nabla C_i$, where the diffusion coefficient $D_i=(l_1/l_i) D$ is assumed to be inversely proportional to the protein length following the Stokes-Einstein equation. Alternatively, the membrane diffusivity can be influenced by the membrane anchors, but our results are fairly insensitive to the molecular diffusion term (see SI) and we ignore them here. The third term on the right side is a drift in response to membrane deformation at a rate $\frac{k_bT D_i}{\mu}\nabla(C_i\nabla h(h-l_i)$ \cite{Qi:PNAS2001,Burroughs:BioP2002}, where $k_b T$ is the thermal energy. The last two terms correspond to receptor binding at a rate $(C_{i,0}-C_i)K^{on}$ and unbinding at a rate $C_iK^{off}$. The kinetic rates $K^{on}_i$ and $K^{off}_i$ are described in terms of the mean first passage time over an energy barrier \cite{Bell:1984,Kramer:1940}, with a distribution centered around the natural protein length ($l_i$) and being a function of $l_i/l_2-h$, given by 
\begin{equation}
\begin{split}
\textrm{Bond formation:~}~~&K^{on}(l_i)=K^{on}_i=\frac{1}{\tau_k}\exp\left(-\left(\frac{{\frac{l_i}{l_2}}-\frac{h}{l_2}}{{\frac{\sigma_{on} l_i}{l_2}}}\right)^2 \right)\\
\textrm{Bond depletion:~}~~&K^{off}(l_i)=K^{off}_i=\frac{1}{3 \tau_k}\exp\left(-\left(\frac{\frac{l_i}{l_2}-\frac{h}{l_2}}{\frac{\sigma_{off} l_i}{l_2}}\right)^2 \right),
\end{split}
\label{eq:rate}
\end{equation}
where $\tau_k$ is the kinetic time. To favor protein binding for $h\sim l_i$, we assume that proteins lose their bonds three times slower ($3\tau_k$) \cite{Figge:epjD2009} than the rate at which they form. Although the exact form of these rates are not known, if we assume that the off-rate increases with spring tension, so that proteins would unbind as $h\ll l_i$ and $h\gg l_i$ and in its simplest form given by a constant off-rate ($\sigma_{off}=\infty$) in Eq. \ref{eq:rate} (see SI). Experiments show that the the different protein pairs form non-overlapping patterns \cite{Grakoui:science1999,Monks:1998, Kaizuka:2007}, which we mimic in the choice of the width of the kinetic distributions $\sigma_{on}=0.2$ and $\sigma_{off}=0.6$ \cite{Carlson2014}. By narrowing the distributions generate wider protein free areas that separate TCR-pMHC and LFA-ICAM rich regions. In contrast, increasing the distribution widths make the different protein species overlap, which is unrealistic. We focus here on protein transport due to physicochemical processes driven by protein binding and membrane deformation and have neglected the role of active cytoskeleton dynamics in the cell e.g. polarized release of T-cell-receptor-enriched microvesicles \cite{Choudhuri:2014}, endocytosis and exocytosis \cite{Stinchcombe:Nat2006}. 

\section{Dimensional analysis and scaling laws}

\subsection{Dimensional parameters}
\begin{table}
    \begin{tabular}{l l l}
        \hline
        \textbf{Description} & \textbf{Notation} & \textbf{Reference}~\\ \hline \hline
        Fluid viscosity ~ & $\mu=4 \times 10^{-2}~ Pa\cdot s$ &  \\
        Cell membrane Young's modulus~ & $E= (0.08-80) \times 10^6 Pa$~&  \\ %\hline
        Membrane thickness~ & $b= 8 \times 10^9~m$~&  \\ %\hline
        Poisson ratio~ & $\nu = 0.5$~ & \cite{Simson:BioP1998} \\ %\hline
        Bending modulus ~ & $B_m=\frac{Eb^3}{12(1-\nu^2)}=4.5\times (10^{-21}-10^{-19})$ J~& \cite{Allard:BioP2012,Qi:PNAS2001} \\ %\hline
           ~& ~& \cite{Simson:BioP1998} \\ %\hline
        Protein stiffness (Hookean spring)~& $\kappa = 1.2 \times 10^{-6} N/m$~ & \cite{Qi:PNAS2001,Reister:NJP2011}\\
                   ~& ~& \cite{Burroughs:BioP2002}  \\ %\hline
        Equilibrium number density TCR~& $C_{1,0}=C_0=2\times 10^{14} m^{-2}$~& \cite{Grakoui:science1999}\\
        Equilibrium number density LFA~& $C_{2,0}=2\times C_0=4\times 10^{14} m^{-2}$~& \cite{Grakoui:science1999}\\
        Natural TCR-pMHC length ~ & $l_1=15~nm$ & \cite{Hartman:PNAS09}\\% \hline
        Natural LFA-ICAM length ~ & $l_2=45~nm$ & \cite{Hartman:PNAS09}\\ %\hline
       % Membrane separation scale~ & $h_0=l_G=45 \times 10^{-9} m$ ~& () \\
        Membrane protein diffusion coefficient~ & $D=5\times 10^{-13} m/s^2$~ & \cite{Hsu:2012PLOS1, Favier:Imm2001} \\ %\hline
        Kinetic on-rate~ & $\tau_k=\tau_1=\tau_2= 1.1\times(10^{-5}-10^{-1})s$~& \\ %\hline
        Kinetic off-rate~ & $\tau_{off}^c=\tau_{off}^g=\tau_k/3 $ s~&   \cite{Figge:epjD2009}\\ %\hline
        Cell diameter~ & $L= 10~\mu m$~ & \cite{Grakoui:science1999} \\
        Hydrodynamic time scale & $\tau_{\mu}=\frac{\mu}{C_0 \kappa l_2}= 3.7 \times 10^{-3}$ s& \\
        Thermal energy~&$k_bT=4.34\times 10^{-21} J$ & \\ 
        Distribution width on-rate & $\sigma_{on}=0.2$& \\
        Distribution width off-rate & $\sigma_{off}=0.6$& \\
        Pressure scaling & $p_0={C_0\kappa l_2}=10.8~Pa$&\\

        \hline \hline
    \end{tabular}
     \caption{Description of the material parameters that appear in Eq. 1-4.\label{si:para}}
\end{table}
%\clearpage

The material properties of the cell, the fluid and the proteins that are relevant to the IS and needed as input into Eq. 1-4 are summarized in Table \ref{si:para} as reported in previous work in the literature. 

\subsection{Dimensionless numbers}
It is natural to scale the horizontal length scales using the cell size, i.e. $[x,y]\sim L$, the height of the synaptic cleft using the typical protein length i.e. $h\sim l_2$, and the pressure by the local receptor force/area, i.e. $p\sim C_0\kappa l_2\equiv p_0$, and time by a viscous time, i.e. $\tau_{\mu}=\frac{\mu}{C_0\kappa l_2}$.

In Eq. \ref{eq:p}-\ref{eq:rate}, the use of the scaled variables $p(x,y,t)=p^*(x,y,t) p_0=p^*(x,y,t){C_0\kappa l_2}$, $h(x,y,t)={h(x,y,t)^*}{l_2}$, $x={x^*}{L}$, $y={y^*}{L}$, $t={t^*}{\tau_{\mu}}$, $C_i(x,y,t)={C_i(x,y,t)^*}{C_0}$ yields six non-dimensional numbers  that govern the dynamics of protein patterning, as shown in Table \ref{tab:nd}. They are: $B=\frac{B_m}{\kappa C_0 L^4}$ which describes the ratio of pressure generated by membrane bending and the protein spring pressure, $l_1/l_2$ is the relative ratio between the natural length of the proteins, $l_2/L$ is the aspect ratio of the membrane gap, $Pe=\frac{L^2C_0\kappa l_2}{D\mu}$ is the ratio between advection and diffusion, $M=\frac{D \mu}{k_bT C_0 l_2}$ is the ratio between protein diffusion and protein sliding mobility, $\tau=\frac{\tau_{\mu}}{\tau_k}=\frac{\mu}{\tau_k C_0 \kappa l_2}$ is the ratio between the local hydrodynamic time $\tau_{\mu}$ and the kinetic time  $\tau_k$ (Table \ref{si:para} ). As we will show, our results are insensitive to variations in $Pe$, $M$ and initial conditions (see SI), and only the dimensionless numbers $B$ and $\tau$ control the qualitative aspects of our phase space of patterns. 

The two important dimensionless numbers $B$ and $\tau$ can described the potential variations in the membrane properties and/or the protein biochemistry across different experiments. In particular, the membrane properties depends on its composition, where the presence of inclusions e.g. cholesterol, peptides, proteins, can alter its stiffness. $\tau$ depends on the fluid in the synaptic cleft and the biochemistry of protein binding. In particular, if $\tau>1$ bonds form rapidly relative to the time for fluid flow in the cleft which is then rate limiting, and conversely when $\tau<1$, fluid flow is fast relative to bond formation which is then rate limiting. 

\begin{table*}
\centering
    \begin{tabular}{l l}
        \hline
        \textbf{Description} & \textbf{Non-dimensional number} \\ \hline \hline
        Membrane bending/protein stretching ~ & $B=\frac{Eb^3}{12(1-\nu^2)\kappa C_0 L^4}=\frac{B_m}{\kappa C_0 L^4}=2\times(10^{-7}-10^{-9})$\\
        Aspect ratio membrane height/length~ & $\frac{l_2}{L}=4.5\times 10^{-3}$ \\ %\hline
        Protein aspect ratio TCR-pMHC/LFA-ICAM& $\frac{l_1}{l_2}=\frac{l_1}{l_2}=\frac{1}{3}$\\ %\hline
        Diffusive/advective time scale~ & $Pe=\frac{L^2C_0\kappa l_2}{D\mu}=5\times 10^{4}$ \\ %\hline
        Protein sliding mobility/protein diffusion~ & $M=\frac{k_bT C_0 l_2}{D \mu}=2.0$\\
        Hydrodynamic/kinetic time scale ~ & $\tau=\frac{\tau_{\mu}}{\tau_k}=\frac{\mu}{\tau_k C_0 \kappa l_2}=3\times(10^{-3}-10)$\\  \hline \hline
      %  Protein advection & $\frac{l_C h_0}{L^2}=6.75\times 10^{-6}$ \\ \hline \hline
    \end{tabular}
\caption{By substituting the scaled variables in Eq. \ref{eq:p}-\ref{eq:rate}; $p(x,y,t)=p^*(x,y,t) p_0=p^*(x,y,t){C_0\kappa l_2}$, $h(x,y,t)={h(x,y,t)^*}{l_2}$, $x={x^*}{L}$, $y={y^*}{L}$, $t={t^*}{\tau_{\mu}}$, $C_i(x,y,t)={C_i(x,y,t)^*}{C_0}$ gives the non-dimensional numbers above. These non-dimensional numbers characterize the relative influence of membrane mechanics, protein kinetics, geometry and hydrodynamics.\label{tab:nd}}              
\end{table*}

 \begin{figure}[!ht]
 \centering
 \includegraphics[width=1.0\linewidth]{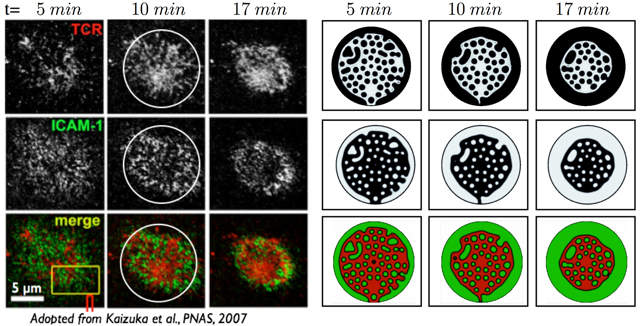}

\caption{Comparison between an experimental (left) and numerical (right) realization of the TCR-pMHC and LFA-ICAM protein patterning in the IS. The simulations are based on Eq. \ref{eq:p}-\ref{eq:rate} allowing fluid flux at the edge of the IS, where the height and number of proteins per membrane area is fixed. The protein pattern is determined by varying $B$ and $\tau$. In the experiment a T-cell interacts with an antigen seeded lipid bilayer \cite{Kaizuka:2007}. The upper row shows the density of bonded TCR-pMHC, the middle row the bonded LFA-ICAM proteins and the last row their merged RGB-channel. The right panel shows the numerical simulation with $B=2\times 10^{-9}$ and $\tau=15$ where the dimensional times correspond to the non-dimensional times $(\frac{h_0}{L})^2\times t=[1.7,~3.3,~5.7]$. All other non-dimensional numbers are reported in Table \ref{tab:nd}. The white circle illustrates the numerical domain and the experiment and simulation are compared at the same instances in time. At short-times, protein clusters nucleate on the membrane, with a dynamics given by the interplay between membrane mechanics, protein kinetics and fluid flow. At late times protein clusters interact and coalesce into large spatial patterns that mimic pSMAC and cSMAC structures. A ''donut shaped`` LFA ring surrounds a dense TCR region at the center of the synapse at late times.\label{fig:comp}}
\end{figure}

\begin{figure}[!ht]
%\flushright
%\subfigure[{\normalsize First row: Contour plots of the time history of the pressure ($p$), superimposed with the velocity vectors. Second row: Contour plot of the number of TCR-pMHC (red) and LFA-ICAM (green), see Fig. 3 for color scale.}]{%$(\frac{h_0}{L})^2\times t=7.0$]{
{{\includegraphics[width=0.85\linewidth]{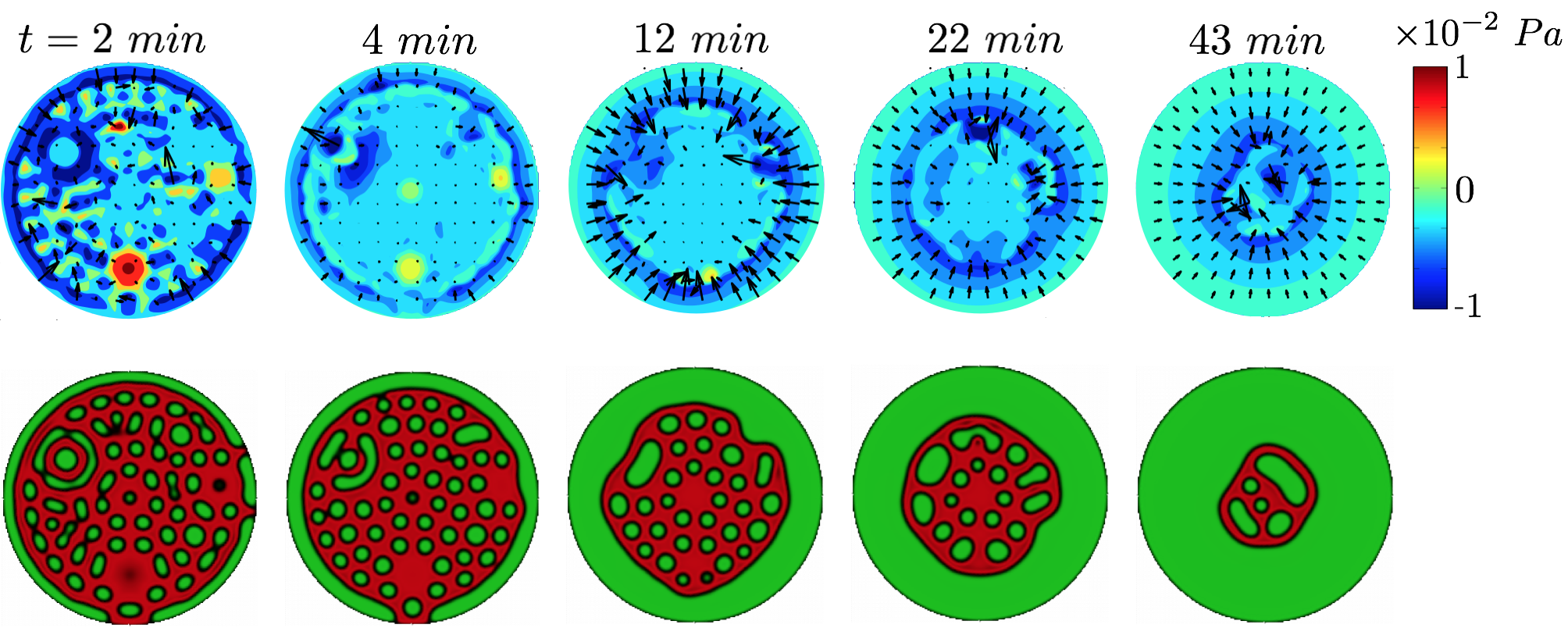}}
}
%\flushright
%\subfigure[{\normalsize Total number of attached TCR-pMHC as a functions of time (t).}]{%$(\frac{h_0}{L})^2\times t=7.0$]{
{{\includegraphics[width=.45\linewidth]{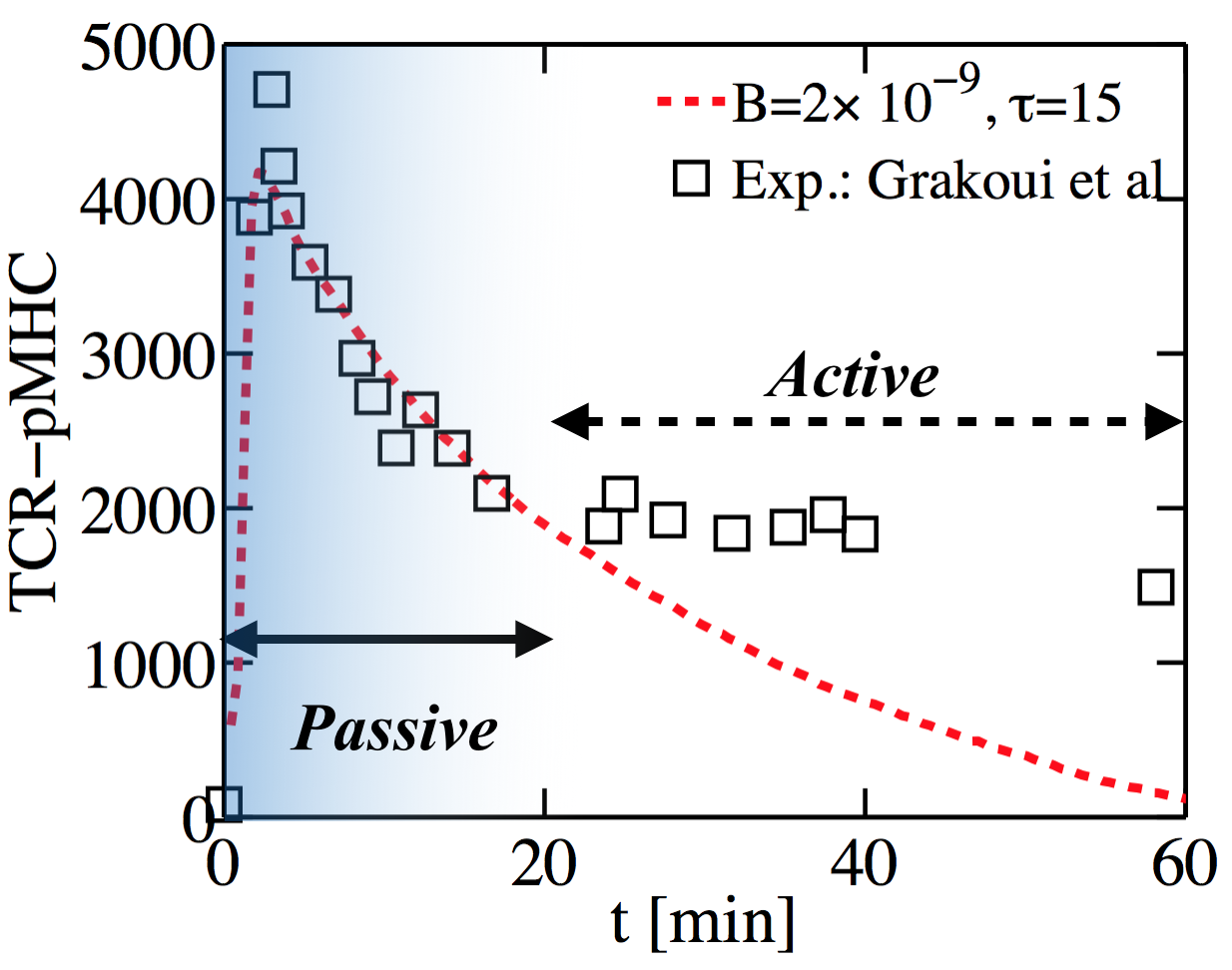}}
}
%\flushright
%\subfigure[{\normalsize Total number of attached LFA-ICAM as a functions of time (t).}]{%$(\frac{h_0}{L})^2\times t=14.0$]{
{{\includegraphics[width=.45\linewidth]{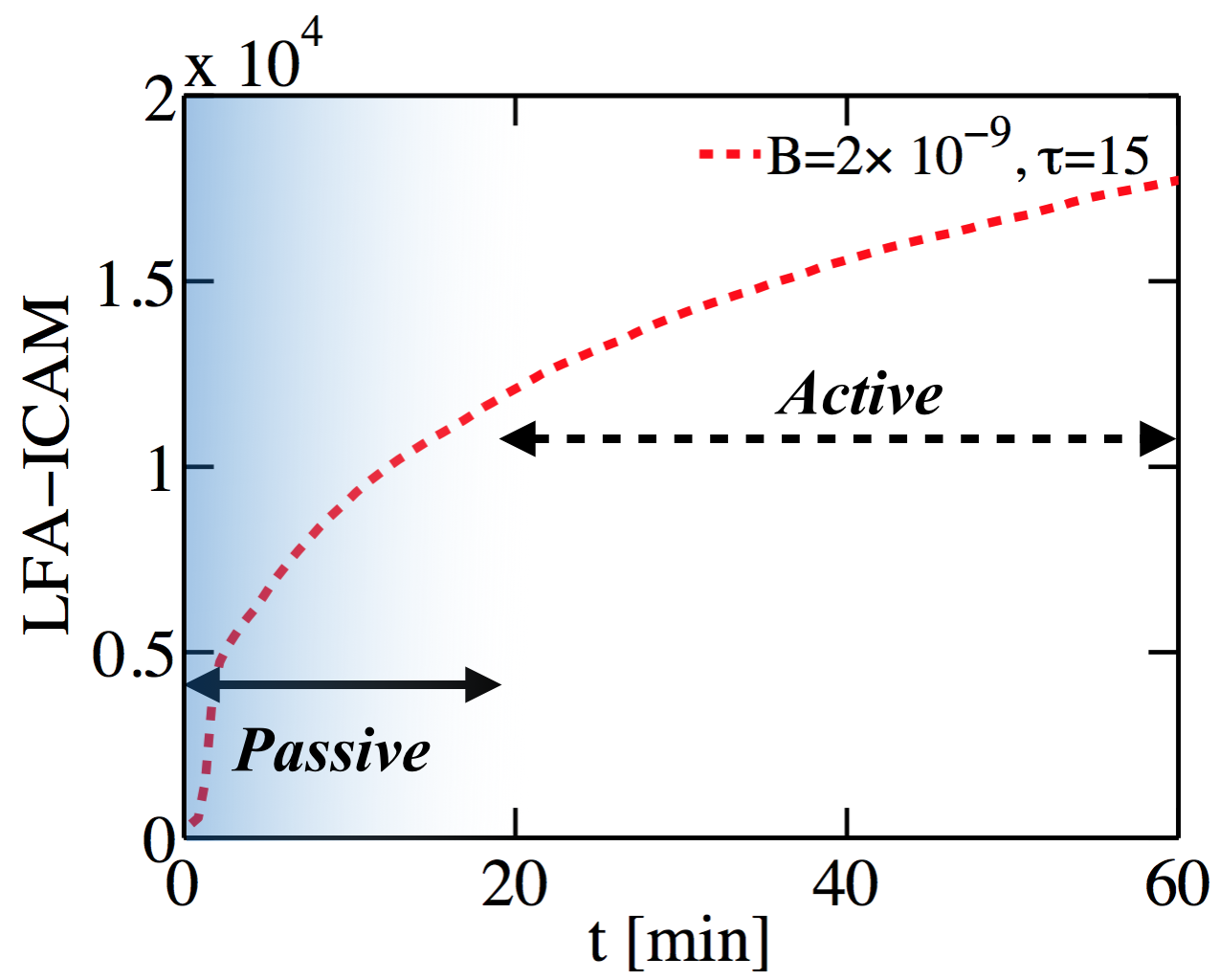}}
}
\put(-220,200){(a)}
\put(-350,-10){(b)}
\put(-100,-10){(c)}

\caption{\small{(a)-(c) Simulation of Eq. \ref{eq:p}-\ref{eq:rate} with $B=2\times 10^{-9}$ and $\tau=15$ and the other dimensionless numbers are reported in Table 2. (a) Contour plots of the time history of the pressure ($p$) and the velocity ($h^2\nabla p$), superimposed with the velocity vectors. The second row shows the corresponding protein pattern of TCR-pMHC and LFA-ICAM, see Fig. 3 for color scale. At short times ($t<4~min$) the nucleation and coalescence of protein domains generates a local flow field. At late times ($t\geq12 min$) a global centripetal flow is generated that "compress`` the TCR cluster radially generating a "bulls-eye"-like protein pattern, which becomes unstable at $t\approx~60 min$. (b-c)The total number of attached receptors (b) TCR-pMHC and (c) LFA-ICAM. (c) Direct comparison between of the total number of attached TCR in the IS in simulation and in the experiment by \cite{Grakoui:science1999} shows that the results are in good agreement for $t<20~min$. This suggests a "waiting-time`` for the active cytoskeleton processes for protein domain organization and that passive dynamics suffices to describe the short-time formation and organization of protein domains. But the long-time IS dynamics and its stability is suggested to be dominated by active processes.\label{fig:recept}}}
\end{figure}

\begin{figure}[!ht]
\centering
\includegraphics[width=0.8\linewidth]{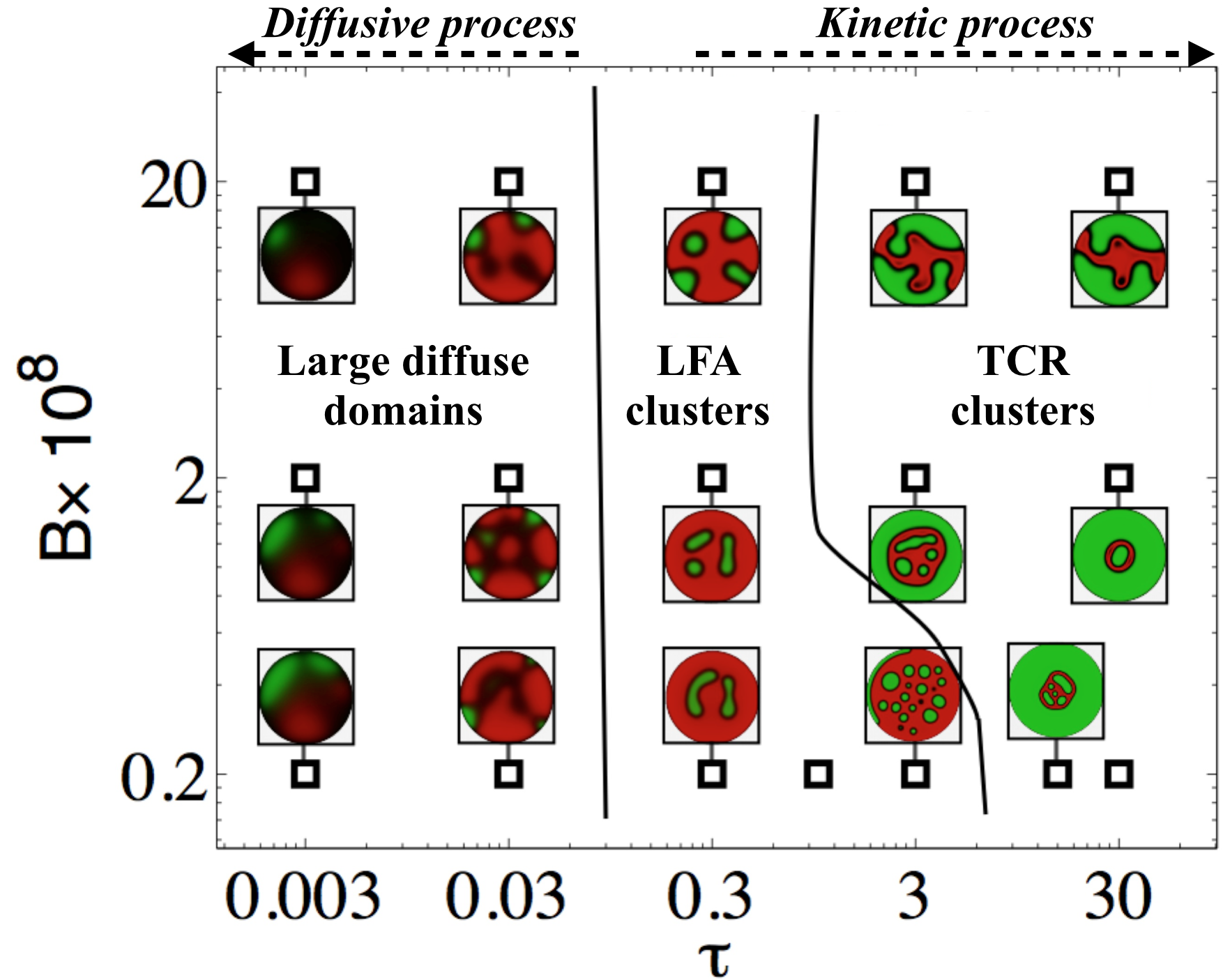}
\caption{Phase space that characterizes the different regimes of membrane protein patterns as a function of $B=\frac{Eb^3}{12(1-\nu^2)\kappa C_0 L^4}$ and $\tau=\frac{\tau_{\mu}}{\tau_k}=\frac{\mu}{\tau_k C_0 \kappa l_2}$ (see Table 2), here presented on logarithmic axes. The simulations are based on  Eq. \ref{eq:p}-\ref{eq:rate} and the patterning is measured at $t=40~min$ where a synaptic pattern is typically formed in experiments \cite{Monks:1998, Grakoui:science1999,Kaizuka:2007,Hartman:PNAS09}, i.e. in dimensionless units $(L/l_2)^2 t= 16$. Two different protein patterns are identified; large diffuse patches and dispersed kinetic clusters, which are categorized into three regimes. In the diffusional dominated limit ($\tau<0.3$) large diffusive patches are predicted that translocate on the membrane. A transition to a dispersed protein pattern is observed for $\tau>0.3$. In the intermediate regime ($0.3\leq\tau\leq3$), long-lived LFA clusters form on the membrane. When the protein dynamics is an active process ($\tau>3$) micro-scale TCR clusters nucleate and coalesce as they are transported radially forming a central dense pattern. In the kinetic regime we see that the cluster size varies as a function of $B$, similar to our scaling prediction $\approx B^{\frac{1}{4}}$. Note that at equilibrium, all simulations predict a flat membrane with a single protein phase for the case where fluid flux at the edge of the IS is free and the membrane height and number of proteins per membrane area is fixed.\label{fig:regime}}
\end{figure}

\subsection{Length scales}
Two characteristic lengths are observed in the IS, the micro-cluster scale $l_c$ and the large domain scale $L$. From Eq. \ref{eq:p} we derive a scaling law for the cluster size, by balancing the spring pressure and bending pressure ${B_m}l_2/{l_c}^4\approx C_0 l_2$ that leads to 
\begin{equation}
{l_c}\approx(\frac{B_m}{C_0\kappa})^{\frac{1}{4}}.
\label{eq:lc}
\end{equation}
For the simulated $B_m$ (SI) the deformation length varies between ${l_c}\approx(\frac{B_m}{C_0\kappa_1})^{\frac{1}{4}}=70-200~nm$ i.e. in dimensionless units $l_c^*=B^{\frac{1}{4}}\approx0.02-0.06$ for $B\in[10^{-9}, 10^{-7}]$, qualitatively consistent with experimental observations \cite{Kaizuka:2007,Hartman:PNAS09}.

\subsection{Time scales}
Protein patterning at the micro-cluster ($l_c$) size occurs on short time scales ($\tau_c$), while patterning at the cell scale ($L$) occurs on long time scales ($\tau_L$). Fluid continuity and force balance embodied in Eq. \ref{eq:tf} yields a short  time scale $\tau_c$  corresponding to drainage on the micro-cluster scale $l_c$, given by
\begin{equation}
\tau_c=12(\frac{l_c}{l_2})^2\tau_{\mu}=12(\frac{B_m}{C_0\kappa l_2^4})^{\frac{1}{2}}\frac{\mu}{C_0\kappa l_2}.
\end{equation}
Substituting in parameter values yields $\tau_c\approx 0.1-1~s$ i.e. in dimensionless time units $\tau_c^*=12(\frac{l_c}{l_2})^2\approx 24-240$ (see SI).  Fluid drainage on the cellular scale $L$  yields a long  time scale given by 
\begin{equation}
\tau_L=12(\frac{L}{l_2})^2\tau_{\mu}=12(\frac{L}{l_2})^2 \frac{\mu}{C_0\kappa l_2}
\end{equation}
Substituting parameter values yields $\tau_L\approx 40~min$ i.e. in dimensionless units $\tau_L^*\approx5\times 10^4$. 

\section{Numerical experiments}
\subsection{Methods}
To solve the nonlinear system of Eq. \ref{eq:p}-\ref{eq:rate}, we numerically discretize these with a finite element method (see SI) in two-dimensions, which gives the membrane topography in three-dimensions. For consistency with experimental observations, the simulations are performed in a circular domain {that capture the central region of the cell-to-cell contact, which is assumed not to be influenced by the motion of the cell  leading edge}. At the edge of the IS the membrane is assumed to be torque free with no bending moment ($\nabla^2 h=0$) and at a constant pressure ($p=0$), which allows fluid flux through the boundary. The membrane is pinned at the edge ($h=0.5 l_2$) and the equilibrium number of proteins per membrane area at that given height ($C_1=C_2=0.01C_0$) see \cite{Carlson2014} and SI for details. The membrane is initialized with six small Gaussian shaped bumps of different widths ($\approx 0.1L$) and amplitude ($(0.075-0.1) l_2$). Additional information about the numerical method \cite{Amberg1999}, \cite{Boyanova2012}, parameter sensitivity and alternative boundary conditions are in the SI.

\section{Results}
Within the phase space of $B$ and $\tau$, we start by considering a cell that has a stiffness that scales with the thermal energy $B_m\approx k_BT$ and binding rates that are similar to those reported in experiments \cite{Grakoui:science1999} $\approx 10^{-4} M s$, with an association constant $\approx 0.1 M^{-1}$ giving $\tau_k\approx 10^{-5}$. We note that the hydrodynamic time scale is larger $\tau_{\mu}\approx 3\times10^{-3}s$ than $\tau_k$ suggesting that the IS dynamics is rate limited by the fluid flow i.e. $\tau\gg 1$, which we verify below.

In Fig. \ref{fig:comp} we show the time evolution of the IS for these parameters ($B=2\times 10^{-9}$, $\tau=15$) and we see that the qualitative behavior of our model is consistent with the observed asymmetric IS dynamics seen multiple times \cite{Grakoui:science1999,Kaizuka:2007,Beemiller:2012,Sims:Cell2007,Brossard:2005} (see Movie 1), and recovers the temporal dynamics and the cluster sizes seen in experiments, associated with the presence of dense non-overlapping regions of TCR-pMHC and LFA-ICAM, which vary with time. At short times dispersed micron-sized protein clusters nucleate on the membrane, with a characteristic cluster size $\approx 1\mu m$ (containing $\approx 160$ proteins). These protein clusters are transported by the centripetal fluid flow generated by membrane deformation. At long times, we see the appearance of larger spatial protein domains, with a "donut-shaped`` LFA-ICAM structure (peripheral SMAC) surrounding a dense central domain of TCR-pMHC (central SMAC) (Fig. \ref{fig:comp}). This similarity is particularly striking since we did not evoke any active processes that are present in a cell.

The two-dimensional simulations of the trans-membrane proteins allow for a direct comparison with the asymmetric IS found experimentally \cite{Grakoui:science1999,Kaizuka:2007,Beemiller:2012,Sims:Cell2007,Brossard:2005}. {To illustrate how these transport processes are correlated with domain coarsening, we show the pressure and velocity fields in Fig. \ref{fig:recept}a. At short times ($t<4~min$) the nucleation and coalescence of protein domains at a length scale $\approx l_c$ generates a local flow field, while at long times ($t>12~min$) the flow occurs over a global length scale $\approx L$ wherein the centripetal flow moves the clusters to the center of the domain and coarsens the protein pattern.} In Fig. \ref{fig:recept}b we directly compare the dynamics of the TCR clusters in the simulation with  experiments \cite{Grakoui:science1999}. With increasing time, the number of attached TCR rapidly increases upon first contact  as micro clusters nucleate. A distinct peak in the number of attached TCR is observed around $t\approx 5~min$ in Fig. \ref{fig:recept}b, followed by a decay in the number of attached receptors over longer times. The agreement with experiments for $t<20~min$ is striking since no active processes are evoked and suggests that the slow dynamics of fluid drainage in the synaptic cleft limits the rate of protein patterning during the early stages of IS dynamics.

At longer times ($t>20~min$) the results of the simulation and experiments deviate from each other, indicating an important role for active processes to stabilize the dynamical synapse. Over this period ($\approx 60~min$), a distinctive feature in the experiment \cite{Grakoui:science1999} is the appearance of  a stable dense circular region  of TCR-pMHC surrounded by a "donut-shaped" ring of LFA-ICAM.  Compared to the TCR, the attached LFA display a different dynamics as they increase monotonically in time (Fig. \ref{fig:recept}c) and around $t\approx 60~min$ saturates the nearly flat membrane. A similar time evolution is also observed in the experiment by \cite{Grakoui:science1999}, but their choice of scaling makes a direct comparison challenging.

Moving beyond the direct comparison with experiments, we turn to a qualitative phase-space of protein patterning characterized by $\tau, B, Pe, M$, initial conditions and boundary conditions. Our simulations show that the pattern dynamics are insensitive to variations in $Pe, M$ and the initial conditions (SI). However, the scaled \emph{membrane stiffness ($B$)} and the \emph{ratio of time scales ($\tau$)} are the main players responsible for variations in the patterns. In Fig. \ref{fig:regime}, we show this in terms of a phase diagram of pattern possibilities illustrated by snapshots of the protein distributions at $t=40~min$, a stage corresponding to a mature IS \cite{Kaizuka:2007, Monks:1998, Lee:2002, Grakoui:science1999}.
 
Two distinct protein patterns may be identified corresponding to either large diffuse domains or a dispersed micro cluster phase. We can further categorize the latter into two distinct regimes. For $\tau<0.1$ the membrane proteins  fail to form an IS and their dynamics are primarily dominated by diffusive fluxes and the results are insensitive to $B$. For $\tau>0.3$ islands of non-overlapping micro-scale protein clusters form different shapes on the membrane. For $0.3\leq\tau\leq 3$ long-lived LFA clusters form at the center and at the edge of the membrane. In this regime, kinetic processes and diffusive fluxes make comparable contributions. By further decreasing the kinetic rate ($\tau>3$) the protein dynamics become hydrodynamically limited with a sharper protein interface. In this regime, a large central domain of TCR with a few internalized LFA micro-clusters form on the membrane, which is surrounded by LFA alike the IS. We emphasize that at very long times the equilibrium state corresponds to a nearly flat membrane adhesively bound by either TCR or LFA to the bilayer. {However, a change in boundary condition that replaces the constant pressure along the edge with a vanishing fluid flux, i.e. $\nabla p\cdot \mathbf{n}=0$ where $\mathbf{n}$ is normal vector at the boundary, leads to an arrested inhomogeneous protein pattern (see SI and Movie 2). However, this late stage regime does not influence the initial nucleation and growth of protein domains}.

Our calculations of the protein patterns show that the formation of a synapse-like protein pattern only occurs in the hydrodynamically limited regime for $\tau>0.3$. In this regime, protein clusters nucleate at short-time $t\approx 1~min$ forming a patchy pattern, with a characteristic cluster size that scales as $l_c\approx\left(\frac{B_m}{C_0\kappa}\right)$ (Eq. \ref{eq:lc}). These micro scale protein clusters move centripetally by the self-generated fluid flow {since membrane deformation by protein binding displaces the interstitial fluid and generates flow}, which assists sorting and formation of protein domains. Cluster translocation leads to self-interactions and the formation of large protein domains at long times $t\approx 30~min$ with the characteristic  "donut-shaped`` LFA domain that surrounds a central domain dens in TCR (see Fig. \ref{fig:regime}), similar in structure to what is often referred to as a peripheral-SMAC and a central-SMAC in experiments \cite{Grakoui:science1999,Kaizuka:2007,Beemiller:2012,Sims:Cell2007,Brossard:2005}.

\section{Discussion}
To get at an accurate description of the spatiotemporal dynamics of protein patterning in the IS we have formulated and solved a minimal mathematical model that account for membrane mechanics, protein binding kinetics and hydrodynamics, while setting the stage for the quantification of passive and active mechanisms in the IS. Our theory captures the length and time scales of protein patterning seen in experiments {by only accounting for the passive processes}. Our scaling laws for the size of protein clusters, as well as short and long time protein patterning dynamics are corroborated in simulations without ad-hoc physical assumptions. In particular we show that slow dynamics of fluid drainage in the synaptic cleft can account for the time scales of protein patterning. Direct comparison of our computations with experiments by \cite{Grakoui:science1999} suggests that at early times  passive dynamics suffices to describe the formation and organization of trans-membrane receptors, and suggests a natural time scale for when active processes come into play. {Our passive model of the immune-cell synaptic cleft is a simplification, where we have neglected the mechanisms by which receptor binding generates signaling that triggers internal activity e.g. actomyosin polymerization, endo-/exo-cytosis, release of TCR through microvesicles, local recruitment of integrins etc.  Since all these effects can influence  the patterning dynamics, to challenge our passive physicochemical theory and to help identify the key biophysical process underlying the formation of the IS, we now turn to some experimentally testable predictions.} 

First, a characteristic spatial scale for membrane deformation is predicted by $l_c=(\frac{B_m}{C_0\kappa})^{\frac{1}{4}}$, where $B_m$ is bending stiffness, $C_0$ protein number density and $\kappa$ protein stiffness. Since $l_c$ is fairly parameter insensitive, modifying cell membrane rigidity (wheat germ agglutinin (WGA) \cite{Evans:PP1985}), the protein number density (corralling \cite{Groves:science1997}) or protein stiffness (linker length \cite{Bird:1988}) would only produce moderate changes in cluster size.

Second, two time scales are derived for the short and long time dynamics. At short time protein clusters nucleate $\tau_{c}\approx(\frac{l_c}{L})^2 \frac{\mu}{C_0\kappa}=(\frac{l_c}{l_2})^2 \tau_{\mu}$ and at long time and length scales large protein domains form $\tau_L\approx(\frac{L}{l_2})^2\tau_{\mu}=(\frac{L}{l_2})^2 \frac{\mu}{C_0\kappa l_2}$. In contrast  to the prediction for $l_c$, both $\tau_{c}$ and $\tau_L$ are sensitive to changes in protein number density ($C_0$), protein ($\kappa$) and membrane stiffness ($B_m$), which can be experimentally changed by corralling, linker-length and WGA and will change these three parameters, respectively. Thus, our theory predicts that the time scales for the IS can be changed, without much variation in the spatial features. 

Third, our numerical simulations predict only protein domains for $\tau_{\mu}<\tau_{k}$, identifying protein kinetics as a critical component in the IS formation. Thus, proteins need to bind faster ($\tau_k$) than the characteristic fluid flow ($\tau_{\mu}$) to form a protein pattern. By changing the adhesion molecules the kinetic rate $\tau_{k}$ can be varied and $\tau_{\mu}$ can be modified by playing with the protein number density (corralling) or protein stiffness (linker length), whereas fluid viscosity is expected to be challenging to alter.

Fourth, the effective boundary condition at the periphery of the synaptic cleft is found to be a key component in the longevity of the pattern. Simulations allowing fluid flux through the edge of the IS show that the SMACs become unstable at long times. The formation of a tyrosine phosphatase network at the synapse periphery generates additional resistance to fluid drainage and may limit the rate of mass flux. Thus, the proteins at the boundary of the IS are predicted at one component that regulate its stability and suggests that a disruption of this protein network would affect its longevity. 

Fifth, the fluid motion in the membrane gap has hitherto not been quantified. Such experiments may be feasible with quantum dot tracing techniques \cite{Derfus2004}) and may shed new light on the fluid pathway during the patterning. Fluid can either become trapped in the inter-membrane gap, internalized by the cell or escape at its edge. Another time scale appears for a flow though a porous cell protein network ($\sim \frac{\mu}{K_p C_0 \kappa l_m^2}$), which depends on its permeability $K_p~[m^{-2}]$ and thickness $l_m~[m]$. 

Sixth, we predict nucleation, translation and sorting of protein clusters in the absence of active processes. Recent observations by \cite{James:2012Nat} of non-immune cells show protein patterning and makes an experimental platform ideal to challenge our spatiotemporal predictions. 

Our mathematical model presented here is a minimal and general theoretical skeleton for an accurate description of this class of cell-to-cell interaction phenomena, which is likely useful beyond the IS and understand broader aspects of cell adhesion, communication and motility.

\section{Acknowledgements}
The computations in this paper were run on the Odyssey cluster supported by the FAS Science Division Research Computing Group at Harvard University

\clearpage

\renewcommand{\theequation}{S.\arabic{equation}}
\renewcommand{\thefigure}{S.\arabic{figure}}
\renewcommand{\thesection}{S.\arabic{section}}
        \setcounter{figure}{0}
        \setcounter{table}{0}

\section{{Supplementary Information for "\emph{Elastohydrodynamics and kinetics of protein patterning in the immunological synapse}" A. Carlson and L. Mahadevan}} 
%\begin{affiliations} \item School of Engineering and Applied Sciences, Kavli Institute for Bionano Science and Technology, and Wyss Institute, Harvard University, Cambridge, USA. \item Departments of Physics, and Organismic and Evolutionary Biology, Harvard University, Cambridge, USA. 
%\end{affiliations}

\subsection{Description of movies}

Movie 1: The dynamics of protein patterning for the case when the fluid flux at the boundary is free to vary, but the pressure is fixed. This causes the pattern to eventually decay.

Movie 2: The dynamics of protein patterning for the case when the fluid flux at the boundary vanishes. This causes the pattern to eventually get arrested.

\subsection{Problem parameters}
\begin{table}[b]
 \caption{Description of the material parameters that appear in Eq.(1-4).}
    \label{si:para}
    \begin{tabular}{l l l}
        \hline
        \textbf{Description} & \textbf{Notation} & \textbf{Reference}~\\ \hline \hline
        Fluid viscosity ~ & $\mu=4 \times 10^{-2}~ Pa\cdot s$ &  \\
        Cell membrane Young's modulus~ & $E= (0.08-80) \times 10^6 Pa$~&  \\ %\hline
        Membrane thickness~ & $b= 8 \times 10^9~m$~&  \\ %\hline
        Poisson ratio~ & $\nu = 0.5$~ & \cite{Simson:BioP1998} \\ %\hline
        Bending modulus ~ & $B_m=\frac{Eb^3}{12(1-\nu^2)}=4.5\times (10^{-21}-10^{-18})$ J~& \cite{Allard:BioP2012,Qi:PNAS2001} \\ %\hline
           ~& ~& \cite{Simson:BioP1998} \\ %\hline
        Protein stiffness (Hookean spring)~& $\kappa = 1.2 \times 10^{-6} N/m$~ & \cite{Qi:PNAS2001,Reister:NJP2011}\\
                   ~& ~& \cite{Burroughs:BioP2002}  \\ %\hline
        Equilibrium number density TCR~& $C_{1,0}=C_0=2\times 10^{14} m^{-2}$~& \cite{Grakoui:science1999}\\
        Equilibrium number density LFA~& $C_{2,0}=2\times C_0=4\times 10^{14} m^{-2}$~& \cite{Grakoui:science1999}\\
        Natural TCR-pMHC length ~ & $l_1=15~nm$ & \cite{Hartman:PNAS09}\\% \hline
        Natural LFA-ICAM length ~ & $l_2=45~nm$ & \cite{Hartman:PNAS09}\\ %\hline
       % Membrane separation scale~ & $h_0=l_G=45 \times 10^{-9} m$ ~& () \\
        Membrane protein diffusion coefficient~ & $D=5\times 10^{-13} m/s^2$~ & \cite{Hsu:2012PLOS1, Favier:Imm2001} \\ %\hline
        Kinetic on-rate~ & $\tau_k=\tau_1=\tau_2= 1.1\times(10^{-5}-10^{-1})s$~& \\ %\hline
        Kinetic off-rate~ & $\tau_{off}^c=\tau_{off}^g=\tau_k/3 $ s~&   \cite{Figge:epjD2009}\\ %\hline
        Cell diameter~ & $L= 10~\mu m$~ & \cite{Grakoui:science1999} \\
        Hydrodynamic time scale & $\tau_{\mu}=\frac{\mu}{C_0 \kappa l_2}= 3.7 \times 10^{-3}$ s& \\
        Thermal energy~&$k_bT=4.34\times 10^{-21} J$ & \\ 
        Distribution width on-rate & $\sigma_{on}=0.2$& \\
        Distribution width off-rate & $\sigma_{off}=0.6$& \\
        Pressure scaling & $p_0={C_0\kappa l_2}=10.8~Pa$&\\

        \hline \hline
    \end{tabular}
\end{table}
%\clearpage
Table \ref{si:para} summarizes material properties that are relevant to the IS synapse, as reported in previous work in the literature and are used as inputs to Eq. 1-4.
%\subsection*{Membrane tension}

\subsection{Boundary conditions}
The boundary condition at the edge of the IS critically affects the final protein pattern since it reflects the role of different biophysical processes associated with membrane deformation, fluid flow and the number of proteins per membrane area. Three different types of boundary conditions for the membrane edge can be prescribed, as given below:
\begin{enumerate}
\item [ ] \emph{Pinned}:~~$\nabla^2 h=0, h=constant, C_1=C_2=constant$ 
\item [ ] \emph{Clamped}:~~$\nabla h\cdot \mathbf{n}=0, h=constant, C_1=C_2=constant$
\item [ ] \emph{Free}:~~$\nabla^2 h=\nabla^3 h\cdot \mathbf{n}=0, \nabla{\mathbf{J_i}}\cdot \mathbf{n}=0$,
\end{enumerate}
where $\mathbf{n}$ is the boundary normal. In a slight abuse of notation, we denote \emph{Pinned} and \emph{clamped} as corresponding to fixing the density of proteins at a given height at the membrane edge, and further letting the torque vanish or fixing the angle at the edge.  For a \emph{shear and moment free} membrane edge, the edge is free and further we assume that there is no protein flux at the boundary. 

Any of these three set of boundary conditions for the membrane height can be prescribed with the additional boundary conditions for the fluid motion that read
\begin{enumerate}
\item [ ] \emph{Free fluid flux}:~~$p=0$
\item [ ] \emph{No fluid flow ($\mathbf{u}=0$)}:~~$\nabla{p}\cdot \mathbf{n}=0$.
\end{enumerate}
In the case where there are few proteins at the membrane edge, the pressure is prescribed (\emph{free fluid flux}) to allow fluid flow into or out of the membrane gap. If the IS is sealed off by a dense protein network, it is hard for the fluid to escape at the edge and \emph{no fluid flux} is a natural prescription.

In the main text, we have used a pinned membrane with a constant pressure at the edge, allowing mass fluid flux through the boundary of the domain. The use of these boundary conditions is based on experimental observations, where only at late times a protein network surrounds the IS. One implication of using a \emph{pinned} and \emph{free fluid flux} boundary conditions is that the protein pattern does not stabilize, which can however be arrested by prescribing \emph{no fluid flow} at the free boundary.  We note that it is possible to extend the mathematical model to account for a free boundary problem for the location of the edge itself, but we avoid this scenario here as it is not relevant to the dynamics of the IS and generates a significant numerical complication, as a dynamic mesh is needed to track the membrane edge. 

\subsection{Computational methodology}
\label{sm:num}
The governing equations (Eq. 1-4) were solved with the open-source finite element toolbox femLego \cite{Amberg1999}. A first order semi-implicit Euler scheme is used for  time marching and all variables are discretized in space using piecewise linear functions. The non-linearity together with the sixth order derivatives in Eq. 2 makes it challenging  to solve. Therefore, we decompose this into three equations,  for the Laplacian of the height ($\nabla^2 h$), the pressure ($p$) and the height ($h$). These three equations are coupled with two additional equations for the proteins, and are solved simultaneously using a Newton iteration method \cite{Boyanova2012}.

We performed a convergence study of the 1D results  in both time and space; different spatial $\Delta x/L=[0.007, 0.0025, 0.001]$ and temporal $\delta t/\tau_{\mu}=[100, 400,1000]$ resolution show no noticeable change in the results. The 2D results presented here use a mesh size ($\Delta x/L=0.006$) and a time step ($(\frac{l_2}{L})^2  \Delta t^*=0.02$).

\subsection{Dependence of dynamics on ratio of hydrodynamic to kinetic time scale $\tau$}

\begin{figure*}[ht!]
%\centering
\begin{overpic}[width=1.0\linewidth]{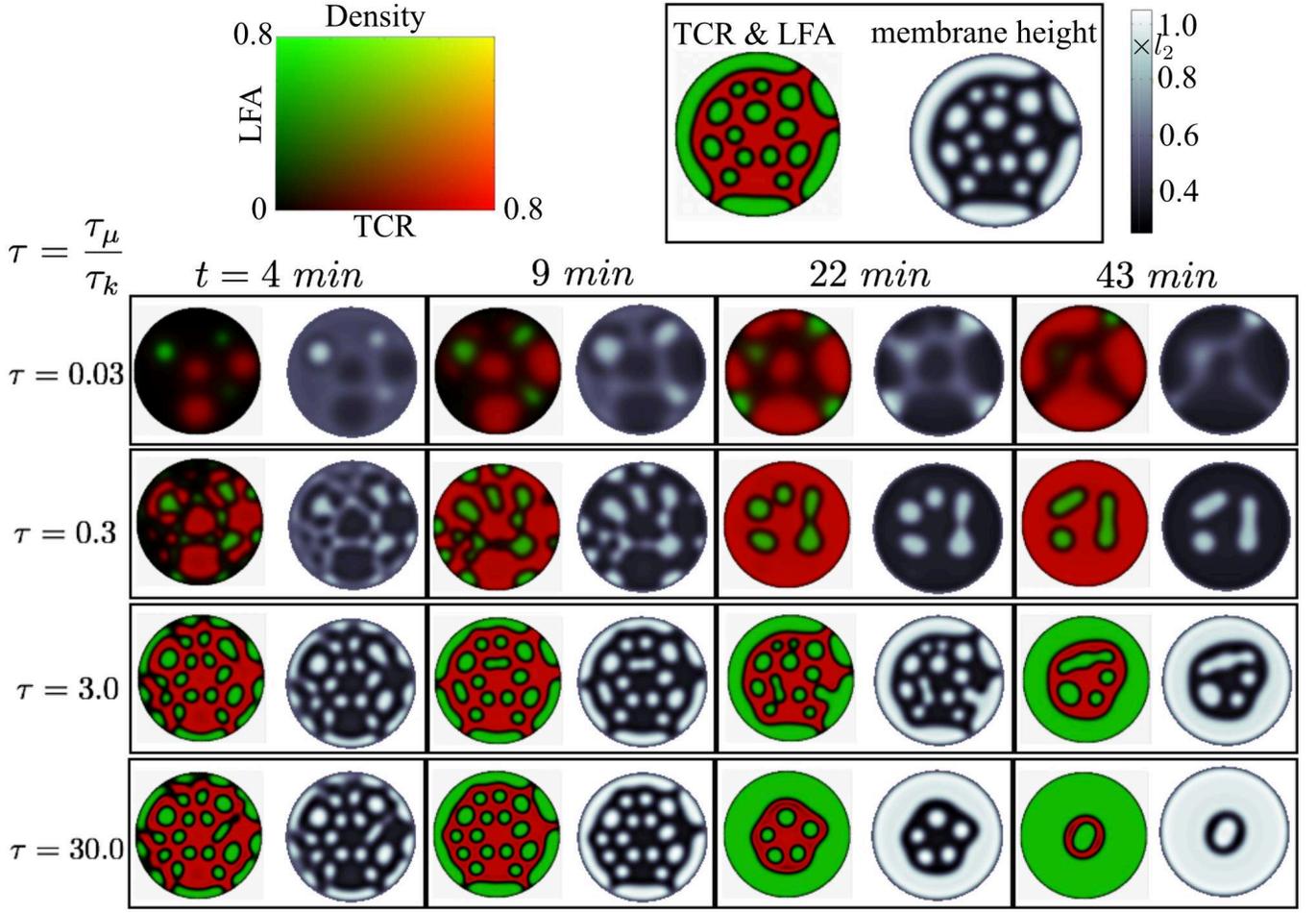}
\put(443,355){\large${ \times l_2}$}
\end{overpic}
\caption{Time history of the attached proteins and the membrane topography as a function of $\tau=[0.03, 0.3, 3.0, 30]$. $\tau=\frac{\tau_{\mu}}{\tau_{k}}=\frac{\tau_{\mu}}{\tau_k}=\frac{\mu}{\tau_k C_0 \kappa l_2}$ is the relative importance between the local viscous time ($\tau_{\mu}$) and the kinetic time rate ($\tau_k$). $B=2\times 10^{-8}$ and the other non-dimensional numbers are given in Table 1. The simulations are based on Eq. 1-4. The color-scale for the density of bonded  LFA (green) and TCR (red) is shown to the upper left corner and the scale bar for the membrane height (black-white) is shown in the upper right corner. For $\tau\ll1$ the dynamics are hydrodynamically limited and no protein clusters are predicted. In contrast, for $\tau>0.3$ clusters of TCR and LFA nucleate at short-time and translocate centripetally at long times forming large protein domains.\label{fig:tau}}
\end{figure*}

To investigate the nature of the spatiotemporal evolution of the trans-membrane proteins we vary $\tau$ while keeping all other parameters fixed (Fig. \ref{fig:tau}). In Fig. \ref{fig:tau} we see that when $\tau \ll 1$ the patterns are kinetically limited. Large protein patches nucleate on the membrane that slowly drift by diffusion. In contrast, {increasing the role of hydrodynamics by the increase in $\tau$} leads to a patchy protein pattern of receptor micro-clusters that are separated by a sharp interface. The clusters move centripetally (see Fig. \ref{fig:tau}), causing the pattern to coarsen as they coalesce, which leads to the formation of large protein domains. At equilibrium the membrane is nearly flat and saturated by a single protein species. 

\subsection{Parameter sensitivity - diffusion, sliding, initial condition, off-rates and membrane tension}
\label{sm:diff}
\begin{figure}[!h]
\includegraphics[width=1.0\linewidth]{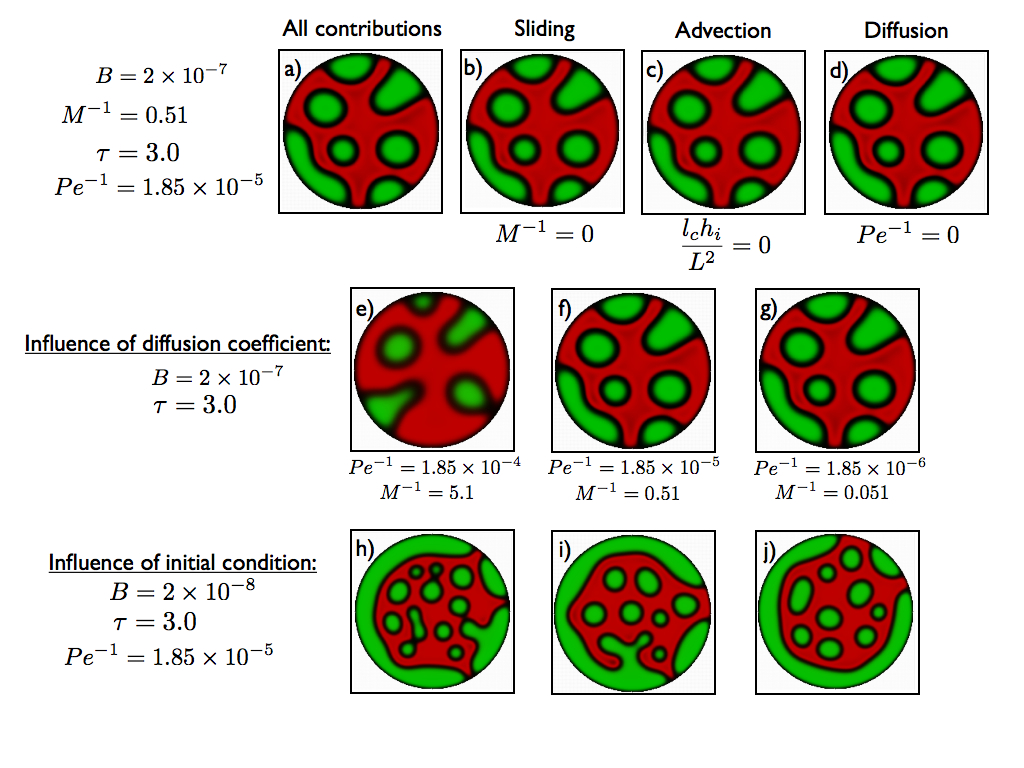}
\caption{Influence of protein diffusion, sliding and advection and the initial condition on the predicted numerical results at time $t=23~min$ i.e. in dimensionless units $\frac{l_2}{L})^2 t^*=7.0$.}
\label{si:diff}
\end{figure}

For given initial conditions, protein diffusion, sliding and advection can influence the dynamics of patterning. To quantify the influence of these properties on the resulting protein patterns, we separately turn off these effects. In Fig. \ref{si:diff} we show the results when  protein sliding is turned off ($M^{-1}=0$), in Fig. \ref{si:diff}b we show the results when  protein advection is turned off ($l_ch_i/L^2=0$), and in Fig. \ref{si:diff}c we show the results when protein diffusion is turned off ($Pe^{-1}=0$) (Fig. \ref{si:diff}d). What is clear from Fig. \ref{si:diff} is that none of these parameters has any significant contributions in the kinetic regime ($\tau=3.0$), where macroscopic patterns persist.

In order to determine the influence of diffusion ($D$) on the resulting dynamics, we varied $D$ over two orders of magnitude (Fig. \ref{si:diff}e-g). By changing the diffusion coefficient the value of both $Pe$ and $M$ change. Increasing $D$ makes the interface between the two boundaries more diffuse, while decreasing $D$ makes the boundaries sharper. Comparing the results for $D=5.0\times 10^{-13}m^2/s$ and $D=5.0\times 10^{-14}m^2/s$ we notice that besides this quantitative feature, the results are indistinguishable and diffusion does not strongly influence the protein patterns over this range. 

To quantify the influence of the initial conditions, we perform simulations with three different initial conditions (Fig. \ref{si:diff}h-j). Initially the membrane has six small Gaussian shaped bumps of different widths ($\approx 0.1 L$), with an amplitude ($(0.075-0.1) l_2$). In the sub-figure to the lower left in Fig. 7 the bumps on the membrane are inverted compared to the simulation to the lower right. The result presented in the middle sub-figure shows a simulation result with an initial membrane shape with six Gaussian bumps at different positions than shown in the left and right sub-figure. Although the detailed shape of the pattern is slightly influenced by the initial condition, the overall dynamics is robust to these changes. 

{We have in this work assumed that the kinetic binding and unbinding rates are described by the means passage time over an energy barrier Eq. 4, which leads to an Gaussian distribution for the on/off rates centered around $l_i$. The off-rates may be a function of the tension in the proteins with a probability of unbinding that increases with the tension up to a given threshold. Eq. 4 generates an effective kinetic rate that takes the form of a double-well, while an off-rate based on the tension in the proteins would remove the two minima and the probability of unbinding approaches a constant value as the proteins are further stretched/compressed. The simplest form of a tension based off-rate is to let $\sigma_{off}=\infty$, where the effective rate ($K_i^{on}-K_i^{off}$) becomes a shift of the gaussian for the on-rate and the probability of unbinding becomes constant for large protein deformation. We have performed additional simulations to verify that our results are not very sensitive to the from of the off-rate, which is demonstrated in the second row in Fig. S.3. Although the detailed shape of the pattern is slightly different, the overall dynamics is robust predicted in the simulation.

Since the membrane has a fluid-like nature, there can also be an influence in the pressure from membrane tension and an additional term $\gamma \nabla^2$ enters into Eq. 1
\begin{equation}
p(x,y,t)={B_m} \nabla^4 h-\gamma \nabla^2 h+\kappa C_1(h-{l_1}) + \kappa \frac{2l_1}{l_2} C_2(h-l_2)
\end{equation}
where $\gamma$ is the membrane tension $[N/m]$. In the tension dominated limit the length scale for membrane deformation scales as $\left(\frac{\kappa C_0}{\gamma}\right)^{\frac{1}{2}}$. Scaling pressure with the characteristic spring pressure $\kappa l_2 C_0$ yields another dimensionless number $\Gamma=\frac{\gamma}{l_2^2 \kappa C_0}$, which is the ratio between pressure from membrane stretching and the pressure from deforming the protein springs. In Fig. S.3 row 3-5 we demonstrate the influence of membrane tension by varying $\Gamma=[10^3 - 10^5]$ e.g. $\gamma=[4-0.04]\times 10^{-3} N/m$ in dimensional units. For $B=2\times 10^{-7}$ we note that as $\Gamma<10^{-5}$ the spatiotemporal dynamics is dominated by membrane bending (Fig. S.3). If the membrane tension is increased, larger protein domains appear $\approx \Gamma^{\frac{1}{2}}$ and if the membrane becomes too stiff the pressure generated by the protein springs is not sufficient to deform the membrane $\Gamma\geq10^{-2}$.
}

\begin{figure}[!h]
\includegraphics[width=1.0\linewidth]{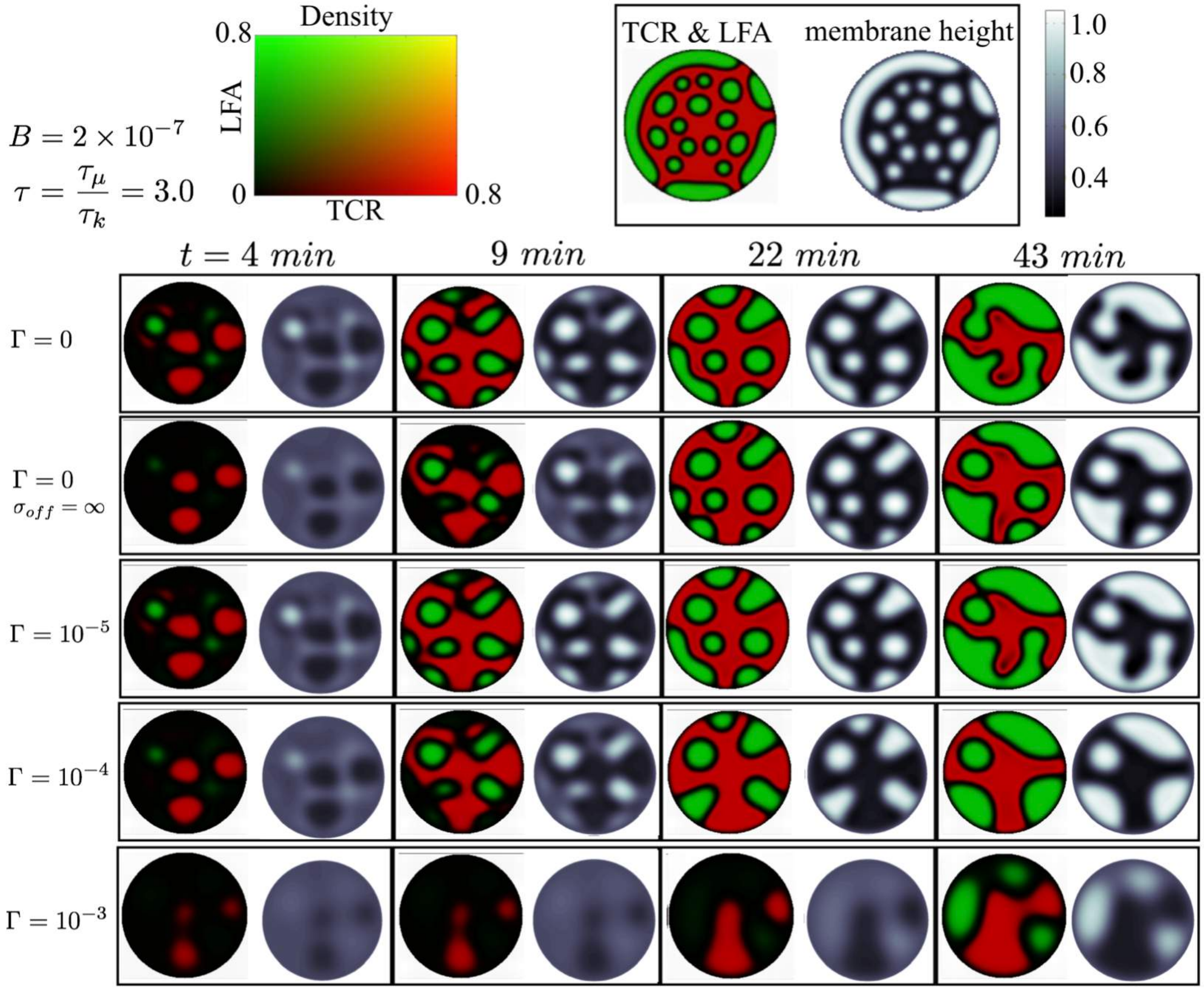}
\caption{Time history of the attached proteins and the membrane topography as a function of off-rate $\sigma_{off}=\infty$ (row 2) and membrane tension $\Gamma=[10^3 - 10^5]$ (row 3-5) for $B=2\times 10^{-7}$, $\tau=3.0$, $Pe=5\times 10^{4}$ and $M=2.0$. $\Gamma=\frac{\gamma}{l_2^2 \kappa C_0}$ is the ratio of pressure from the membrane tension and the protein spring pressure. The color-scale for the density of bonded  LFA (green) and TCR (red) is shown to the upper left corner and the scale bar for the membrane height (black-white) is shown in the upper right corner. These snapshots in time correspond to the dimensionless times $(\frac{l_2}{L})^2\times t^*=[14, 28, 71, 142]$.}
\label{SI:tension}
\end{figure}
\clearpage
\subsection{Sensitivity to boundary conditions}
\label{sm:bcexample}
\begin{figure}[]
\centering
\subfigure[]{%$(\frac{h_0}{L})^2\times t=2.8$ ]{
{\includegraphics[width=0.230\linewidth]{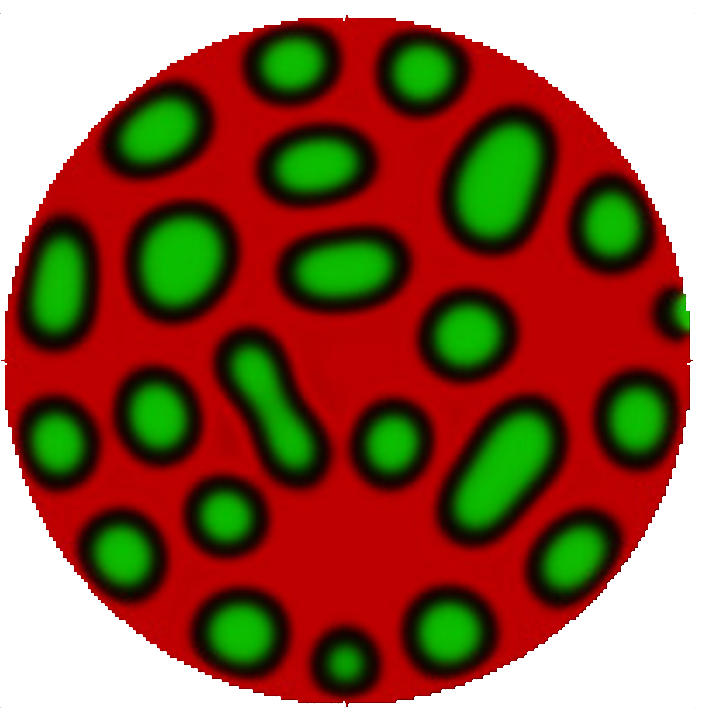}}
}
\centering
\subfigure[]{%$(\frac{h_0}{L})^2\times t=7.0$]{
{\includegraphics[width=0.230\linewidth]{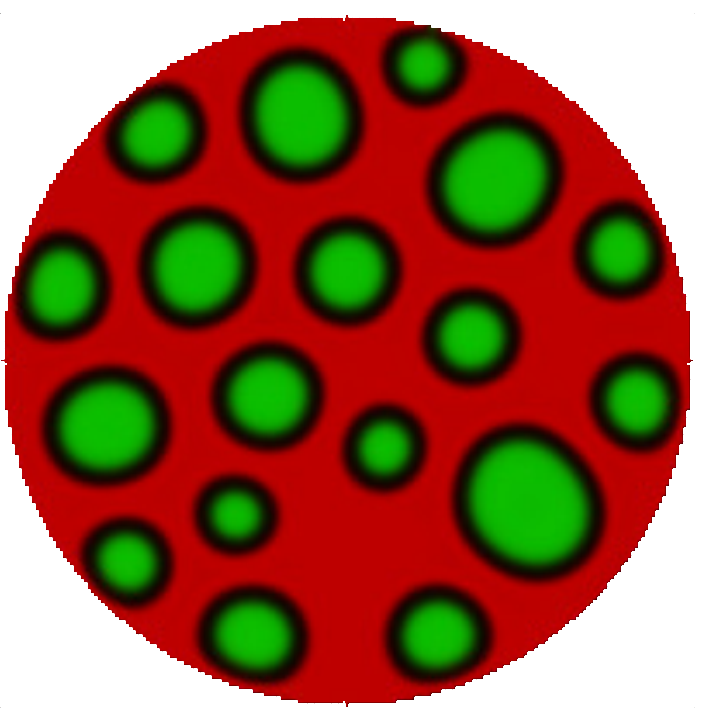}}
}
\centering
\subfigure[]{%$(\frac{h_0}{L})^2\times t=14.0$]{
{\includegraphics[width=0.230\linewidth]{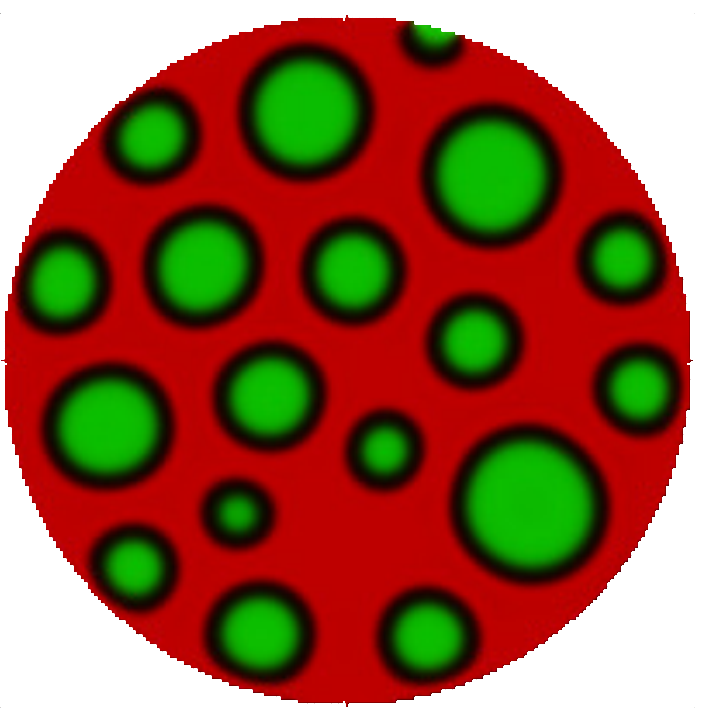}}
}
\centering
\subfigure[]{%$(\frac{h_0}{L})^2\times t=69.0$]{
{\includegraphics[width=0.230\linewidth]{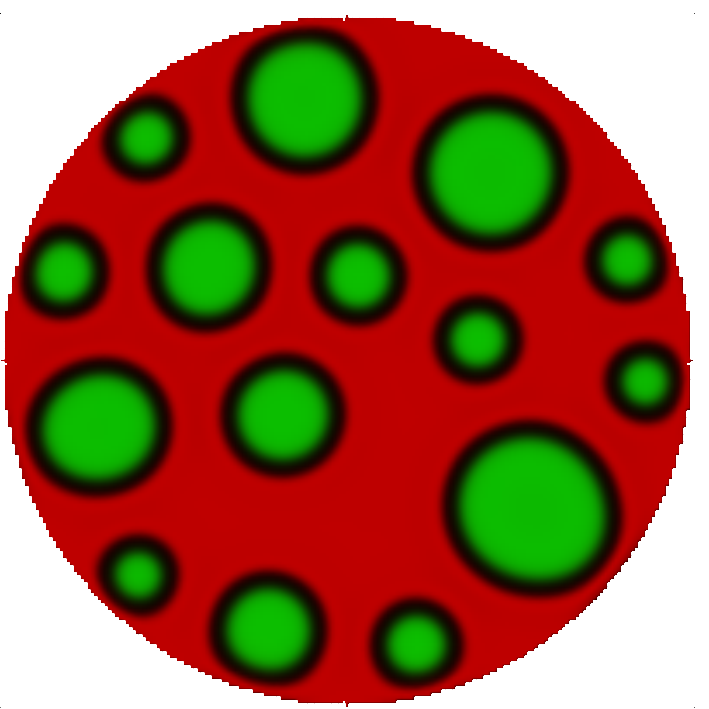}}
}
\caption{Influence of boundary conditions on the IS dynamics ($\tau=3.0, B=2\times 10^{-8}$) at times (a) $t= 10~min$ ($(\frac{l_2}{L})^2 t^*=2.8$), (b) $t=23~min$ ($(\frac{l_2}{L})^2 t^*=7.0$), (c) $t= 47~min$ ($(\frac{l_2}{L})^2 t^*=17$)and (d) $t= 230~min$ ($(\frac{l_2}{L})^2 t^*=69$). At the edge the membrane moves freely using a \emph{shear and moment free} boundary condition, with no fluid flow and a no-flux boundary condition is prescribed for the TCR-pMHC and LFA-ICAM proteins. Contrary to the pinned membrane (Fig. \ref{fig:tau}), which allow in- and out-fluid flow, the protein pattern is arrested at long times.}
\label{sub:BC2D}
\end{figure}

To illustrate how the boundary condition can affects the simulation results in Fig. \ref{sub:BC2D} we show a sequence of snapshots of a simulation with a membrane that is allowed to move freely at the edge (\emph{shear and moment free}) with \emph{no-flux of proteins} and \emph{no fluid flow}. Comparing these results with the case when proteins are free to diffuse through the boundary,  it is clear that the boundary condition affects the protein patterning and serves to arrest the protein pattern at long times.

\end{document}